\documentclass{sig-alternate-10pt}

\usepackage[hyphens]{url}
\usepackage{graphicx}
\usepackage{subcaption}

\usepackage{xcolor}

\usepackage[
  pass,
]{geometry}

\begin{document}

\title{Kadupul: Livin' on the Edge with Virtual Currencies and Time-Locked Puzzles}

\numberofauthors{1}
\author{
       \alignauthor Magnus Skjegstad, Anil Madhavapeddy, Jon Crowcroft\\
       \affaddr{Computer Laboratory}\\
       \affaddr{University of Cambridge}\\
       \email{firstname.lastname@cl.cam.ac.uk}
   }

\maketitle

\begin{abstract}
Devices connected to the Internet today have a wide range of local communication channels available, such as wireless Wifi, Bluetooth or NFC, as well as wired backhaul.  In densely populated areas it is possible to create heterogeneous, multihop communication paths using a combination of these technologies, and often transmit data with lower latency than via a wired Internet connection.  However, the potential for sharing meshed wireless radios in this way has never been realised due to the lack of economic incentives to do so on the part of individual nodes.

In this paper, we explore how virtual currencies such as Bitcoin might be used to provide an end-to-end incentive scheme to convince forwarding nodes that it is profitable to send packets on via the lowest latency mechanism available.  Clients inject a small amount of money to transmit a datagram, and forwarding engines compete to solve a time-locked puzzle that can be claimed by the node that delivers the result in the lowest latency.  This approach naturally extends congestion control techniques to a surge pricing model when available bandwidth is low.  We conclude by discussing several latency-sensitive applications that would benefit for this, such as video streaming and local augmented reality systems. 
\end{abstract}

\section{Introduction}
Devices connected to the Internet today have a wide range of local communication channels available. For example, most new wifi-routers and access points have two or more radios (one for 2.4 GHz and one for 5GHz communication). Connected to each access point there are clients with several radio technologies available, such as Bluetooth and NFC. Other physical communication channels also exist, for example LEDs, cameras~\cite{Scott:2005:UVT:1055959.1055965} and microphones~\cite{madhavapeddy03audio} depending on available hardware.

In urban areas it is possible to create heterogeneous, multihop communication paths using these technologies. As radio waves propagate at the speed of light, these paths offer lower-latency communication. However, there are few economic incentives for edge nodes to act as low-latency data forwarders, and the disincentive of wasting their batteries on other nodes' traffic.

\begin{figure}
    \centering

    \includegraphics[width=0.45\textwidth]{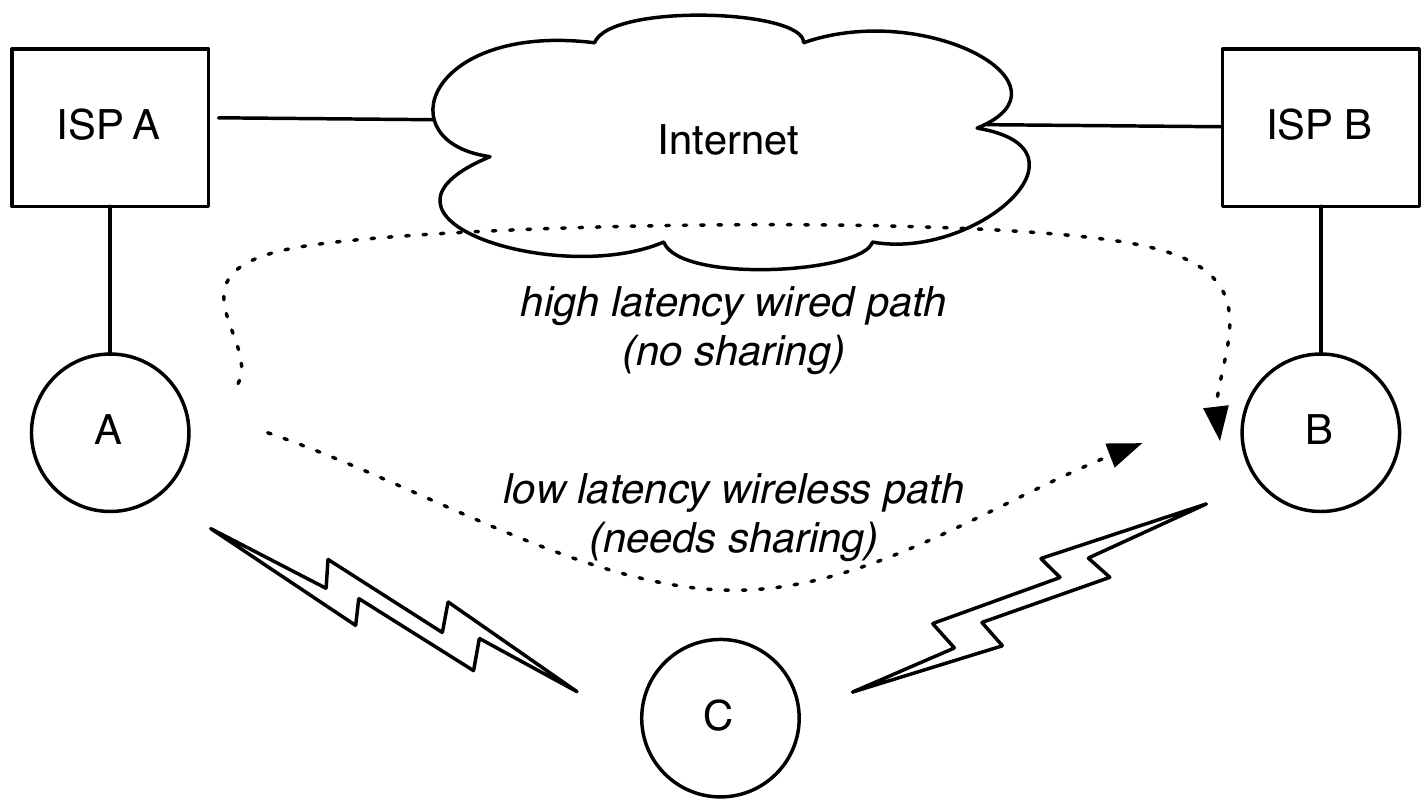}
    \caption{ISP and edge forwarding paths between nodes A and B}

    \label{fig:ispvsedge}
\end{figure}

As an example of how wireless edge nodes can be used for faster forwarding, consider a user A who wants to send messages to user B a few kilometers away, as illustrated in Figure \ref{fig:ispvsedge}. Using a traditional forwarding path the messages may first have to be delivered through the core network of A's ISP (or mobile operator), then be forwarded to the core network of B's ISP, before finally being delivered to user B. The forwarding latency in this example depends on the number of hops and distance to travel via the core networks of the ISPs, not the geographical distance between the nodes. As a result, nodes A and B may experience the same latency whether the geographical distance between them is one or tens of kilometers. 

An alternative forwarding path could be established as a multihop wireless path through intermediate radio devices between A and B, such as node C in Figure \ref{fig:ispvsedge}. Depending on the geographical distance, this forwarding path could achieve significantly lower latency than traditional methods, as well as being resilient to wide-area networking failures since it only depends on the local communications network. There are however few devices today that are willing or able to participate in the network as low-latency edge forwarders. We argue that the primary reason for this is not technical, but caused by lack of incentives compared to the increased workload and need for investment in the edge nodes. For example, it is not uncommon for edge nodes to participate as forwarders in city-wide mesh networks due to bad or expensive Internet connectivity, as in Athens~\cite{athensmesh}, or to act as forwarders to improve communication during a political crisis. Examples of the latter are the Occupy Wall Street movement mesh network~\cite{occupymesh} and the Open Mesh Project in Egypt~\cite{egyptmesh}. In these cases the incentives caused by external factors outweigh the forwarding cost.

The perceived cost of forwarding a message depends on the workload imposed on the edge node. Depending on the technology, this workload may be low (forward on regular Wifi), but one could imagine higher workloads -- such as for a mobile phone that has to turn on Bluetooth discovery for long time periods. The owner of the node may also have to install custom software or invest in upgraded hardware to forward messages faster. Other factors, such as power consumption may also play an important role.

A forwarding incentive can be created by simply offering payment to the forwarders, for example by using a decentralized virtual currency like Bitcoin~\cite{bitcoin}. A useful feature of many virtual currencies is that micropayments can be issued with minimal transaction costs and without relying on a centralized authority. It is more difficult to create a payment system with incentives for minimizing latency, since it is difficult to measure latency objectively in a way that can be accepted by all nodes. For example, the latency observed by the sender, a forwarder and the final recipient may not be the same, and all parties have economic incentives to under- or overreport the latency. 

In this paper we present {\em Kadupul}, a system that incentivises low-latency forwarding between edge nodes without relying on latency measurements. This is accomplished by creating a reward system based on time-locked puzzles~\cite{rivest96}. Time-locked puzzles~\cite{gwern14} can be used to hide information until the puzzle is solved or the solution is provided by the creator or a third-party. Recently, a time-locked puzzle based on Bitcoin was proposed and implemented~\cite{todd14}, which allows Bitcoin rewards to be locked for a given time period and be collected by the first node that solves the puzzle (or is told the solution). 

We build on Todd's time-lock puzzle mechanism~\cite{todd14} to propose a forwarding model for rewarding forwarders by giving them an advantage in solving a puzzle. A forwarder can collect a reward if it provides the correct solution to a puzzle protecting it. Each forwarder is provided with a solution to a reward after it has forwarded a message. The catch is that \emph{each puzzle is public and solvable by anyone after a known amount of time}. This creates a race to forward the message before the puzzle has been solved by other nodes. We also provide examples of applications were low-latency edge node forwarding can be useful, such as in video streaming and augmented reality.

Note that although this paper primarily discusses Bitcoin as a reward system, the ideas described here can be used with other virtual currencies as well as long as they provide a similar underlying P2P protocol.

We will now describe the basic mechanisms in Kadupul and propose several forwarding models based on time locked puzzles as incentives (\S\ref{sec:design}). We then provide examples of how the proposed forwarding mechanisms can be deployed in applications (\S\ref{sec:apps}) and finally discuss future directions (\S\ref{sec:conclusion}).

\section{Design}
\label{sec:design}

We describe the core functionality of Kadupul by first discussing how the forwarding paths are established and negotiated. We then describe in detail how time locked puzzles can be used as a low-latency forwarding incentive and propose several forwarding models based on the mechanism. Examples of applications using these models are provided in Section \ref{sec:apps}. 
Note that a message forwarded by Kadupul can be of any size. For example, a high quality video can be transferred as a single message. 

\subsection{Establishing forwarding paths}
Some coordination must be performed in advance to form heterogeneous multihop communication paths. The forwarders along the path may be able to discover each other directly using radio or other techniques, but if a wide range of technologies are being used over large geographical areas this may not always be possible. 

We assume that in most cases the forwarding path is established by the sender. This requires an initial discovery step where the sender finds potential forwarders that together can forward information from the sender to the receiver. Potential forwarders can for example be discovered based on their location and local communication range, using a decentralized, Internet-based P2P network, such as proposed in for example \cite{p2pdsa} or \cite{geokad}. When a good forwarder is found, the sender contacts the node directly over the Internet and starts a negotiation process.

To negotiate forwarding, the sender asks the potential forwarding nodes it has discovered what their capabilities and forwarding costs are. The capabilities describes the technologies the node has at its disposal, e.g. radio type, range and most importantly, the expected forwarding latency. The forwarding cost is the price of forwarding in a virtual currency such as Bitcoin. 

If the price and range of a forwarding technology is acceptable, the sender attempts to find other nodes that are able to receive the message when it is forwarded. The negotiation process is repeated with these nodes. If no recipients are found, the sender must try to use a different node or different communication techniques. 

The exact discovery and negotiation mechanisms used is out of scope of this paper.

\subsection{Time-locked forwarding rewards}
\label{sec:puzzles}

After establishing the forwarding path, the sender must publish a set of rewards. We now describe the mechanism and protocol used to publish, as well as collect rewards for the forwarding nodes in more detail.

Todd proposes an interesting scheme for implementing time-locked puzzles with Bitcoin rewards~\cite{todd14}. A chain of rewards can be hidden in puzzles and included publicly in the Bitcoin block chain. Each puzzle has a value in Bitcoins that can be collected by any node that knows the solution to the puzzle. The solution can either be provided or can be calculated after a known amount of time. The puzzle is constructed in such a way that it \emph{can be created in parallel, but only solved in serial}.

More specifically, the scheme from \cite{todd14} uses multiple rounds of a SHA256~\cite{rfc6234} hash to calculate a key from a randomly chosen initialization vector for each block of the reward puzzle chain. If the reward chain has for example 10 blocks, 10 initialization vectors are chosen and SHA256 is executed iteratively on each vector to create the keys to unlock the reward. The number of iterations determines how hard it is to recover the keys and how long it will take to unlock the reward without knowing a key in advance.

When the reward chain is made public, the initialization vector in each block (except the first) is obfuscated by XORing it with the accumulated hash result from the previous block. Thus, to decode a reward block without knowing the key you would first need the final hash from the previous block to recover the correct initialization vector. An important feature of the scheme is that the key from the previous reward block has to be revealed publicly in the Bitcoin block chain to claim the reward, forcing each reward collector to reveal the part of the secret they have discovered and thus enabling the recovery of the next initialization vector.

This mechanism can be adapted to create forwarding incentives in several ways, and we discuss four such forwarding models next. Note that control traffic (but not the actual data) is transferred over a higher latency control plane, which for example could be the Internet. 

\subsubsection{Double incentive forwarding}

\label{sec:double}

\begin{figure}
    \centering
    \caption{Negotiation, forwarding and reward collection with double incentive forwarding.}

    \begin{subfigure}[b]{0.45\textwidth}
        \includegraphics[width=\textwidth]{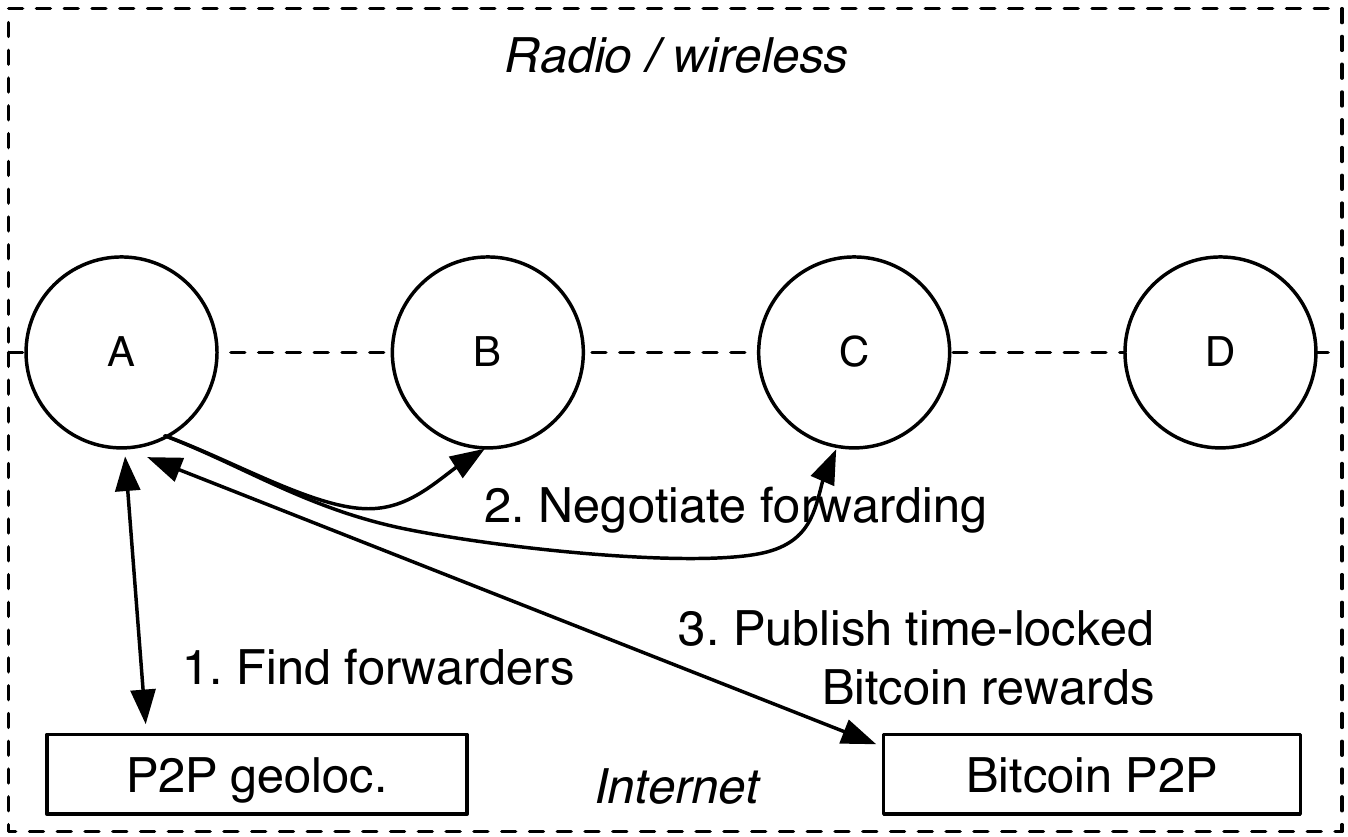}
        \caption{Sender prepares forwarding path}
        \label{fig:path}
    \end{subfigure}

    \begin{subfigure}[b]{0.45\textwidth}
        \includegraphics[width=\textwidth]{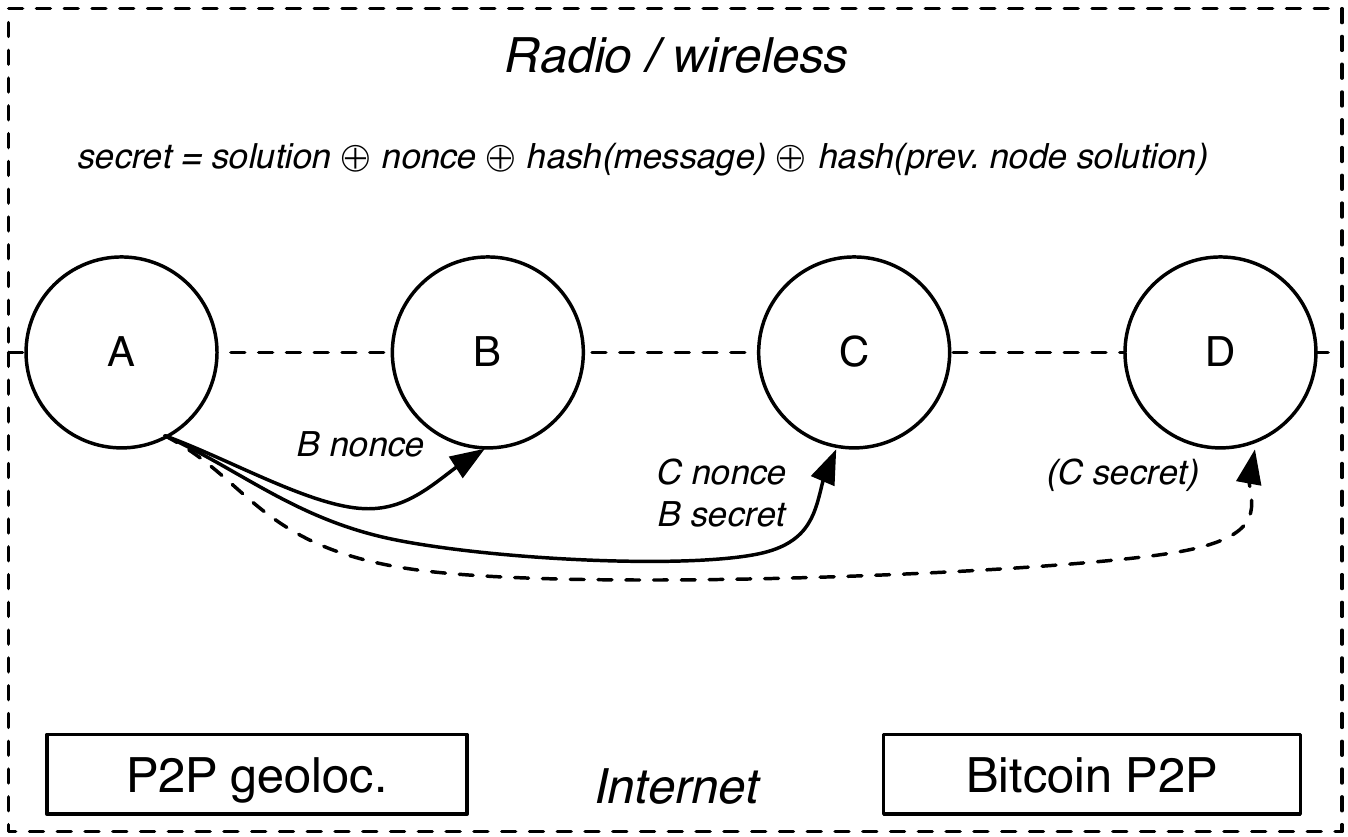}
        \caption{Sender distributes secrets and nonces}
        \label{fig:secrets}
    \end{subfigure}

    \begin{subfigure}[b]{0.45\textwidth}
        \includegraphics[width=\textwidth]{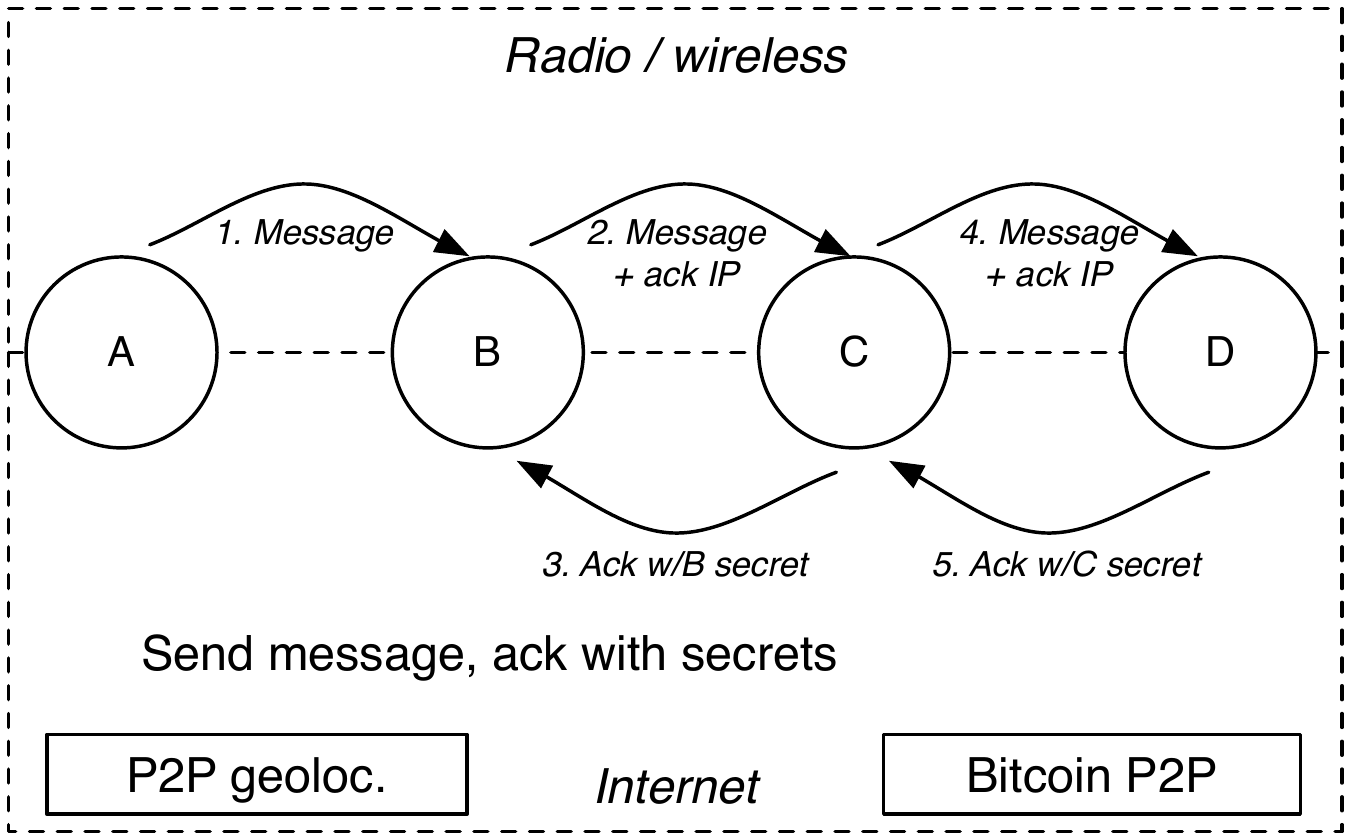}
        \caption{Sender sends message, nodes acknowledge with secrets}
        \label{fig:send}
    \end{subfigure}

    \begin{subfigure}[b]{0.45\textwidth}
        \includegraphics[width=\textwidth]{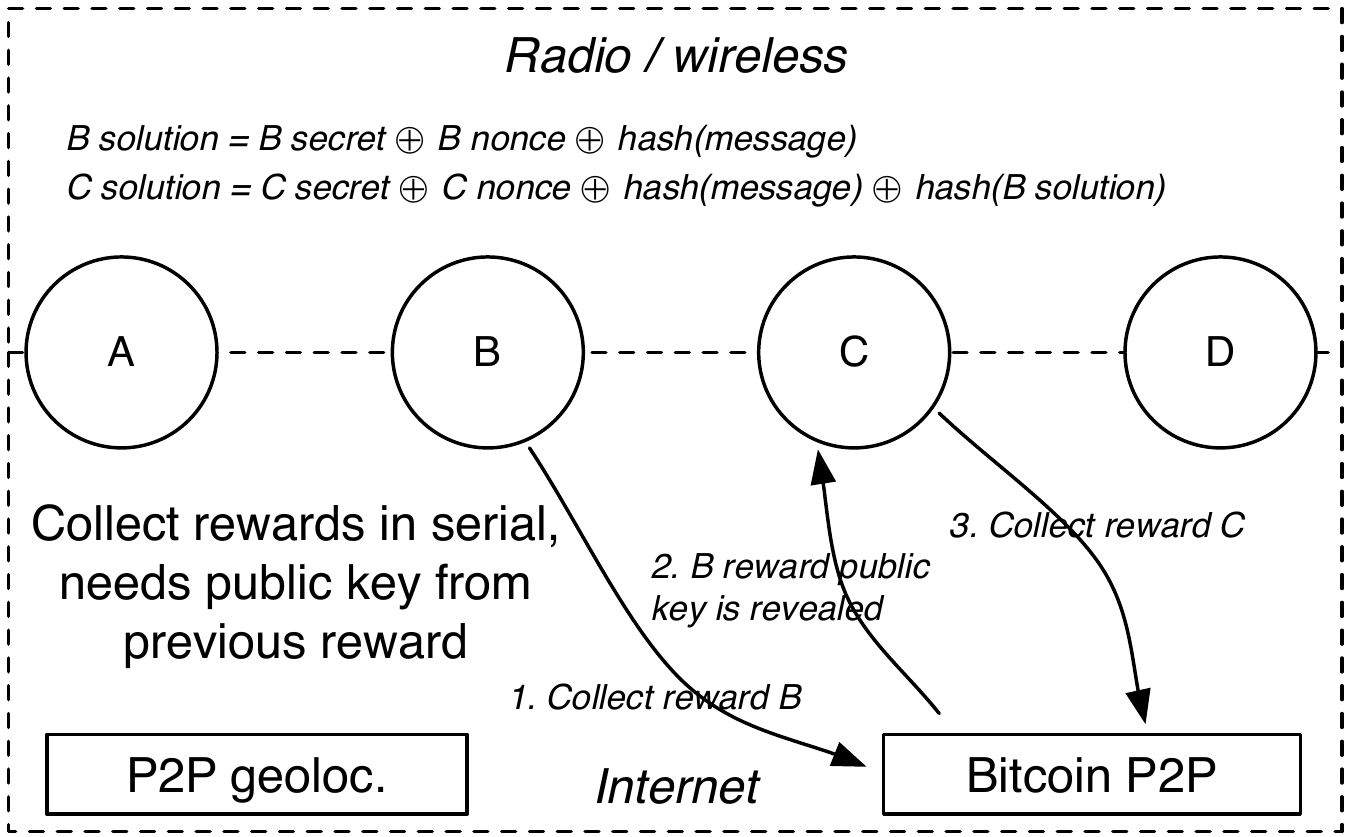}
        \caption{Forwarders reconstruct solutions and collect rewards}
        \label{fig:collect}
    \end{subfigure}

    \label{fig:process}

\end{figure}

The objective in this model is to create a mechanism that makes the forwarders lose their reward unless they forward the message intact to the next hop as soon as possible, but also to create an incentive for assisting other forwarders. The full process is shown in Figure~\ref{fig:process}.

Figure \ref{fig:path} gives an overview of the initial tasks that must be performed before forwarding can begin. First, the sender must find the forwarders and negotiate the forwarding fees. The sender then generates a chain of rewards using time-locked encryption which are published in the Bitcoin block chain. For simplicity, we assume that one reward block is generated for each forwarder. The reward attached to each block may differ in value depending on the terms that were negotiated with the respective forwarder. 

As shown in Figure \ref{fig:secrets}, the sender proceeds to send two values to each forwarder. The first value is a secret that enables the previous hop on the forwarding path to retrieve its reward. The second value is a nonce that when combined with a secret from the next hop on the forwarding path, as well as the hash of the message, results in the key required to unlock one of the rewards.

To avoid having to store the full message in each node, the nodes may use a rolling hash function to hash the message. The nodes then only need to maintain a buffer that contains a window with enough information to create the hash and to act as a message queue for forwarding. 

Now the forwarding itself can begin, as shown in Figure \ref{fig:send}. The message is forwarded along the path and acknowledged by the next hop by sending the secret back to the forwarder. The acknowledgement address in the control plane is also sent with the message - for example an IP address and port number.

Figure \ref{fig:collect} illustrates the reward collection process. After a node has forwarded the full message and received the required secret from the next node, it can reconstruct the puzzle solution and collect a reward. A forwarder will only be able to claim a reward when:

\begin{enumerate}
    \item The previous node in the routing path was able to claim their reward and thus revealed the key of the previous block
    \item It has received the full message successfully so that it can generate a hash
    \item The next node in the path has revealed the necessary secret generated by the sender, thus acknowledging that the full message was forwarded.
\end{enumerate} 

This mechanism ensures that all nodes have incentives both to forward the message to the next hop (to obtain the missing secret to unlock the reward) and to supply the previous hop with its secret (otherwise they are unable to decode their own reward).

As the reward has a time-lock, the nodes are also given an incentive to perform the forwarding as fast as they can, or the encryption may be broken by other nodes in the Bitcoin network trying to claim the reward by brute forcing the keys. The node that can most effectively balance high-throughput and low-enough latency forwarding stands to profit most from solving the majority of puzzles it sees and claiming the rewards.

\subsubsection{All or nothing}
\label{sec:allornothing}
Note that in the forwarding model discussed in Section \ref{sec:double}, the sender and the receiver are never in direct contact in the control plane. The identity of every forwarder must however be revealed to the next forwarder to allow the message to be acknowledged with the correct secret. In a broadcast network this may not be necessary. For example, a forwarder may be instructed to listen to a specific Wifi channel in promiscuous mode to receive the message and then just forward the message as it was received over another broadcast link. It may not be possible for the forwarders to reply with an acknowledgment in the same way if it has a weaker radio than the previous hop.

\begin{figure}
    \centering
        \caption{Broadcast forwarding without revealing forwarder identity to other forwarders.}
        \includegraphics[width=0.45\textwidth]{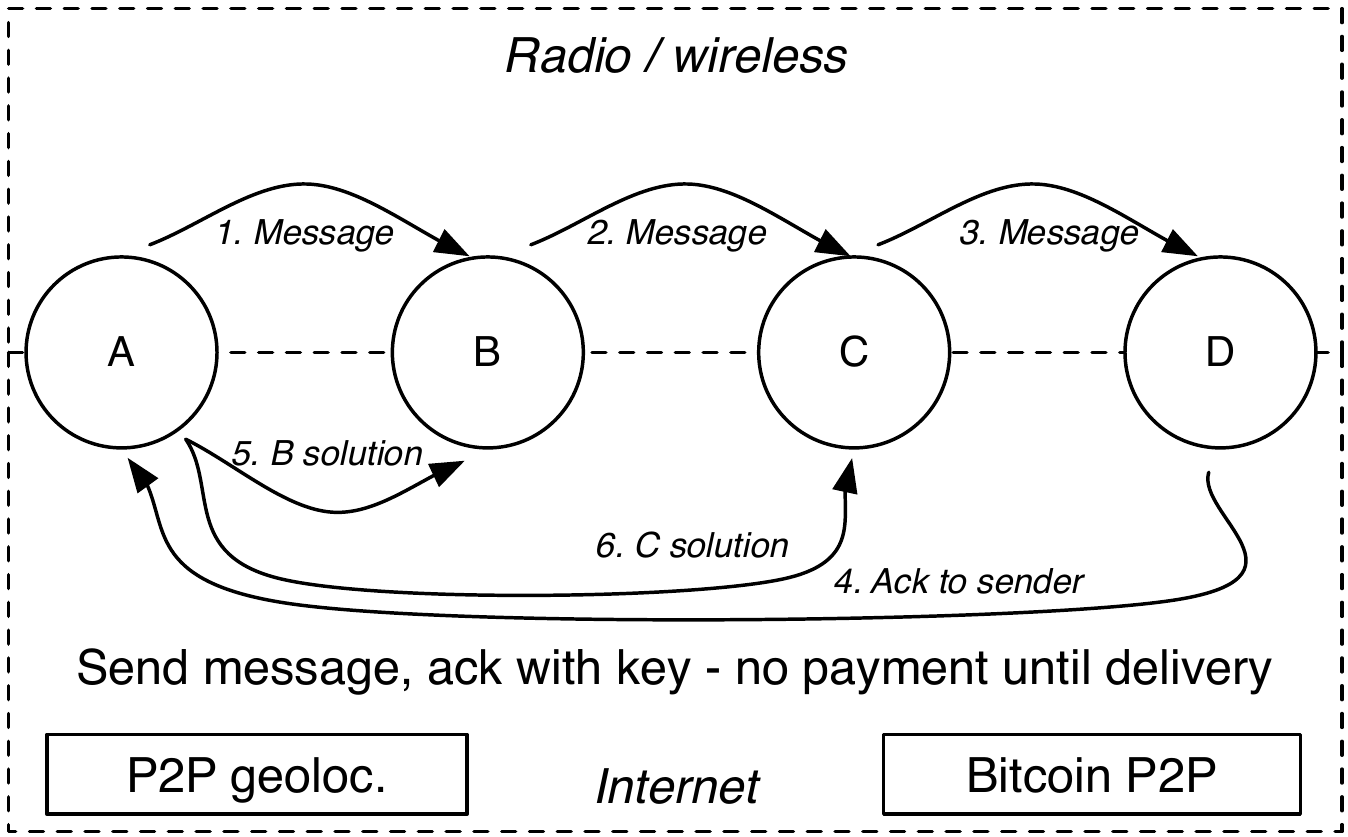}
        \label{fig:process_alt}
\end{figure}

It is possible to use an alternative acknowledgement mechanism to avoid revealing the identity of the forwarders if the sender and the receiver are allowed to communicate over the control plane. This forwarding scheme is shown in Figure \ref{fig:process_alt} and assumes a forwarding mechanism that can hide the Internet address of the forwarder, for example by using anonymous broadcast~\cite{179188}. Instead of distributing secrets in advance, the final receiver acknowledges the receipt of the message to the sender and the sender then unlocks the puzzles for all the forwarders. This scheme requires contact between sender and receiver, but makes it more difficult for the forwarders to collude. It also increases the forwarding risk as none of the nodes will receive their reward if the message is lost or delayed along the way, which may affect the forwarding price. 

This forwarding model can be useful in ``off-the-grid'' mesh networks, where one of the goals is to avoid eavesdropping of traffic by global passive sniffing of conventional networks. 

\subsubsection{Contract forwarding}
Kadupul forwarding may also be used without establishing the forwarding path in advance. In this case the sender negotiates a forwarding contract with another node to bring the message to the recipient. It is then up to the node that accepted the forwarding contract to deliver the message as fast as possible. This node may use any number of subcontractors for the message to reach its final destination. This forwarding model is potentially simpler to implement and use for the sender, but the sender is no longer in control of the path the message travels. It may also increase the forwarding price as more work is left to the forwarders. 

This forwarding model can especially be useful in combination with the other forwarding models. For example, a \emph{pull} based delivery system can be constructed by letting the recipient negotiate a contract with the sender. When the contract has been accepted, the sender uses another forwarding model to deliver the content to the recipient.

\subsubsection{Competing forwarders}
\label{sec:competing}
The ``all or nothing'' model (\S\ref{sec:allornothing}) can be extended to create a competition between forwarders along multiple paths. This can be accomplished by using Random Linear Network Coding~\cite{ho06} (RLNC) or fountain codes~\cite{byers98} to encode partial messages and then forwarding the messages along multiple paths at the same time. RLNC and fountain codes are useful because the original message can be reconstructed when enough coded packets have been received. Fountain codes are end-to-end, but RLNC also allows recoding at intermediate nodes. When the recipient has received enough partial messages to reconstruct the original message, rewards are distributed to the forwarders depending on how much \emph{innovative information} they forwarded. 

This forwarding model can be especially useful for transferring messages that should be distributed to multiple edge nodes in the same area, such as for multicasting video streams.

\subsubsection{Edge node caching}
\label{sec:edgecache}
Kadupul naturally creates incentives for edge nodes to cache content. If an edge node is able to store content that is delivered frequently, it can volunteer to deliver the full content during the negotiation process and collect the full reward for the delivery. It would also be able to deliver the content in much shorter time than if it would have to be forwarded again, allowing it to provide a better offer than its competitors.  Since each node must pay for its own cost of storing the cached data in the hope of future requests, they also have an incentive to develop efficient prediction and cache eviction algorithms.

\section{Applications}
\label{sec:apps}

In the following we give three examples of systems where Kadupul could provide incentives for lowering the latency.

\subsection{Video streaming}
Video streaming is latency sensitive as the video must arrive at least as fast as it is being watched. For end-to-end streaming, such as when watching a popular movie from a video streaming service, Kadupul could provide incentives for the content to be cached in nearby edge nodes, as discussed in Section \ref{sec:edgecache}. If there are many viewers of the same video stream (multicast), the competing forwarders model could be used to distribute the content efficiently over multiple paths, as discussed in Section \ref{sec:competing}.

Live video streams could be requested by using the contract model to pull the content from the sender. The recipient agrees on a contract for delivery by the sender, which then proceeds to set up a double incentive forwarding path towards the final recipient. If there are multiple viewers in an area, the competing forwarder model could also be used. 

\subsection{Off-the-grid forwarding}
The double incentive forwarding model can be used to establish low latency forwarding paths in mesh networks, such as in off-the-grid networks. These paths can be useful for larger data transfers, such as for \emph{streaming video}, or for latency sensitive applications, such as \emph{voice communication}. As an alternative to performing the negotiations over the Internet, the mesh itself could be used as a control plane to establish and negotiate the path. Kadupul forwarding also adds incentives for more nodes to join the mesh network, potentially increasing the overall connectivity. 

A problem with Bitcoin and similar virtual currencies that maintain a public ledger however, is that although the identity of the participants is hidden, all transaction are recorded and published openly. The transaction history of each Bitcoin can be followed and monitored. This could potentially enable third parties to discover the identity of nodes along the communication paths, as well as who paid for the communication and who received it - even years after it occurred. 

In addition to causing privacy issues for the sender and the receiver, this could motivate attacks on common forwarders if their identity becomes known. For example, denial of service attacks could be used in an attempt to stop them from collecting their rewards in time or they could be broken into to steal the keys to their rewards. These problems should however correct themselves over time, as the nodes would be likely to either stop forwarding or not be selected for forwarding in the future (e.g. due to high delivery failure rate).

To mitigate some of the anonymity issues, Bitcoin users have created ``mixing pools'' that enable them to swap Bitcoins with other users. This method is problematic as the users have to trust that mixing pool operator to return their money and that the Bitcoins are exchanged randomly. An extension to Bitcoin called Zerocoin~\cite{zerocoin} has been proposed that enables coins to be mixed using the Bitcoin network itself, but at a considerable resource cost. An improved extension called Zerocash~\cite{zerocash} not only anonymizes the history of each Bitcoin but also the transactions and their amounts, but also has much lower resource overhead than its predecessor. Zerocoin is already being adopted in alternative digital currencies, such as AnonCoin~\cite{anoncoin} and NXTcoin~\cite{nxtcoin}. The inventors of Zerocash have announced that they will release their own implementation in 2014. 

It is thus not unlikely that it will be possible to issue untraceable, anonymous transactions in the near future using digital currencies.

\subsection{Reality Torrent}
\label{sec:reality}
Augmented reality applications are particularly dependent on low latency, since information has to be downloaded and overlayed as a video stream fast for the human brain to not to notice that the rendered information is lagging behind.  Existing systems such as Google Glass use traditional techniques to make a roundtrip to a networked service, whereas Kadupul offers a much more real-time (and battery-friendly) alternative.

In many cases, the AR information can be downloaded in advance and presented when it is needed. In other cases, information must be retrieved from the surrounding area. For example, when entering a building the information about the objects closest to you should be immediately available. As the forwarding path is shorter, a Kadupul forwarding scheme can provide much lower latencies than traditional forwarding in such situations.

When Kadupul is used for augmented reality, the user could pay for low latency access to surrounding content. Another (and perhaps more likely) scenario is that information providers pay for low latency content delivery to users that they care about. For example, users could pay a guided tour company for low latency access to local contextual information in a city they are visiting. This type of forwarding fits with the competing forwarder model (\S\ref{sec:competing}), where the messages being transferred contain local, contextual information.

Edge nodes with storage capabilities may also choose to cache popular content, as the information is geographically dependent and likely to be requested again in the same area. The user would experience a torrent of information that is relevant to the surroundings and the forwarders would be paid by how much new information they provided.

\section{Discussion and Conclusions}
\label{sec:conclusion}

We have proposed a mechanism for creating an economic incentive for low-latency data forwarding and described how the mechanism can be useful for establishing low-latency forwarding paths between edge nodes on the Internet. As examples we have discussed how video streaming, mesh networks and augmented reality applications can benefit from this type of forwarding. We have mainly focussed on edge networks in this paper, but the mechanism can also be used in other types of networks where very low latency should be rewarded.

Although all of the pieces needed to implement this type of forwarding exists today, some technical challenges remain before it can be fully realised. For instance, it may take a long time to set up the initial forwarding path because several potentially slow tasks must be performed before forwarding can begin. For example, the reward puzzles must be generated in parallel, which at least takes the same time as the longest timeout of the rewards. In addition, if Bitcoin is used, the transactions are relatively slow and it takes a while for rewards to propagate in the block chain. This would require that the rewards do not expire until they have had time to be distributed. Today, this makes the method mostly suitable for applications that will use the path for longer time periods.

However, we are confident that these delays will become much smaller in the near future. As virtual currencies are becoming more popular, their protocols and software implementations are constantly being optimized to reduce delays. Furthermore, the number of reward puzzles that can be generated in parallel can dramatically increase if they are calculated using GPUs or FPGAs. Multiple rewards can be given to each hop while still being required to be solved in serial, and so the number of rewards created in parallel can exceed the total number of hops.

Of the application areas we have discussed, the most exciting is the prospect of enabling low-latency access to environmental information for the new generation of wireless headsup display hardware (\S\ref{sec:reality}).  This is an application that, to fall below the threshold perception of the human eye, {\em requires} local wireless connectivity due to the speed-of-light restrictions on making round-trips to remote services, as is done by the majority of existing systems. Kadupul helps to balance the economic needs of service providers and users to deliver a viable model for deploying reliable multihop edge networks, as well as enhancing the resilience of the global network by only using local links when possible.

\section{Acknowledgements}

We would like to thank Marcelo Bagnulo Braun, Thomas Gazagnaire, Andrius
Aucinas, Pedro Ramos Pintos, Richard Mortier and Jeremy Yallop for helpful
discussions of the ideas in this paper. The research leading
to these results has received funding from the European Union's Seventh
Framework Programme FP7/2007-2013 under Trilogy 2 project, grant agreement no
317756.

\bibliography{biblio,rfc}
\bibliographystyle{plain}

\end{document}